\documentclass[twocolumn,english,reprint]{revtex4}
\usepackage[T1]{fontenc}
\usepackage[latin9]{inputenc}
\usepackage{textcomp}
\usepackage{amsmath}
\usepackage{graphicx}

\makeatletter
\@ifundefined{textcolor}{}
{%
 \definecolor{BLACK}{gray}{0}
 \definecolor{WHITE}{gray}{1}
 \definecolor{RED}{rgb}{1,0,0}
 \definecolor{GREEN}{rgb}{0,1,0}
 \definecolor{BLUE}{rgb}{0,0,1}
 \definecolor{CYAN}{cmyk}{1,0,0,0}
 \definecolor{MAGENTA}{cmyk}{0,1,0,0}
 \definecolor{YELLOW}{cmyk}{0,0,1,0}
}

\makeatother

\makeatother

\usepackage{babel}

\makeatother

\usepackage{babel}

\makeatother

\usepackage{babel}

\usepackage{babel}

\makeatother

\usepackage{babel}
\begin{document}

\title{Bending strain-tunable magnetic anisotropy in Co$_{2}$FeAl Heusler
thin film on Kapton\textregistered{}}

\author{M. Gueye$^{1}$, B. M. Wague$^{1}$, F. Zighem$^{1}$, M. Belmeguenai$^{1}$,
M. S. Gabor$^{2}$, T. Petrisor jr$^{2}$, C. Tiusan$^{2}$, S. Mercone$^{1}$,
and D. Faurie$^{1}$}

\email{zighem@univ-paris13.fr ; faurie@univ-paris13.fr}

\affiliation{$^{1}$Laboratoire des Sciences des Procédés et des Matériaux, CNRS-Université
Paris XIII, Sorbonne Paris Cité, Villetaneuse, France}

\affiliation{$^{2}$ Center for Superconductivity, Spintronics and Surface Science,
Technical University of Cluj-Napoca, Str. Memorandumului No. 28 RO-400114,
Cluj-Napoca, Romania}

\date{July 29$^{th}$ 2014 }
\begin{abstract}
Bending effect on the magnetic anisotropy in 20 nm Co$_{2}$FeAl Heusler
thin film grown on Kapton\textregistered{} has been studied by ferromagnetic
resonance and glued on curved sample carrier with various radii. The
results reported in this letter show that the magnetic anisotropy
is drastically changed in this system by bending the thin films. This
effect is attributed to the interfacial strain transmission from the
substrate to the film and to the magnetoelastic behavior of the Co$_{2}$FeAl
film. Moreover two approaches to determine the in-plane magnetostriction
coefficient of the film, leading to a value that is close to $\lambda^{CFA}=14\times10^{-6}$,
have been proposed.
\end{abstract}


\maketitle
The functional properties of devices on non-planar substrates are
receiving an increasing interest because of new flexible electronics
based-technologies. Magnetic thin films deposited on polymer substrate
show tremendous potentialities in new flexible spintronics based-applications,
such as magnetic sensors adaptable to non-flat surfaces. Indeed, several
studies of giant magnetoresistance (GMR)-based devices, generally
composed of metallic ferromagnetic materials deposited on a polymer
substrate, have been made \cite{Bedoya-Pinto2014,Barrault2012,Donolato2013}.
However, in order to develop flexible spintronic devices, materials
with high spin polarization are highly desirable. Half metallic materials
are known to be ideal candidates as high spin polarization current
sources to realize a very large giant magnetoresistance (GMR) and
to reduce the switching current densities in spin transfer based-devices
according to the Slonczewski model \cite{Slonczewski1996}. Among
half metallic materials, Co-based Heusler alloys \cite{Inomata2008}
have generally high Curie temperature such as Co$_{2}$FeAl \cite{Belmeguenai2013}
in contrast to oxide half metals, and thus are promising for spintronics-based
applications at room temperature. However, the knowledge of the magnetoelastic
properties of this Heusler alloys is poor while they could be submitted
to high strains when integrated in flexible devices. 

Among the overall flexible materials, polyimides are organic materials
with an attractive combination of physical characteristics including
low electrical conductivity, high tensile strength, chemical inertness,
and stability at temperatures as high as 650 K. The most common commercially
available (aromatic) polyimide has the DuPont de Nemours registered
trademark Kapton\textregistered{} \cite{Kapton}. In addition to its
widespread use in the microelectronics industry, Kapton\textregistered{}
(PMDA-ODA) has an excellent thermal and radiation stability as evidenced
by its routine use for vacuum windows at storage-ring sources. 

In the case of flexible sample made of polymers coated by a very thin
layer, a very small bending effort can lead to relatively high stress
in the layer, either compressive if it is at the inside edge either
tensile if at the outside ones. Obviously, it depends on the adhesion
between the layer and the Kapton\textregistered{} substrate that is
generally good even if no buffer layer is deposited \cite{Geandier2010,Djaziri2011}.
In this paper, we will show that the magnetic anisotropy of 20 nm
thick Co$_{2}$FeAl (CFA) film grown on Kapton\textregistered{} is
significantly changed through the magnetoelastic coupling by bending
the sample glued on curved Aluminum blocks of different known radii. 

\begin{figure}
\includegraphics[clip,width=8.5cm]{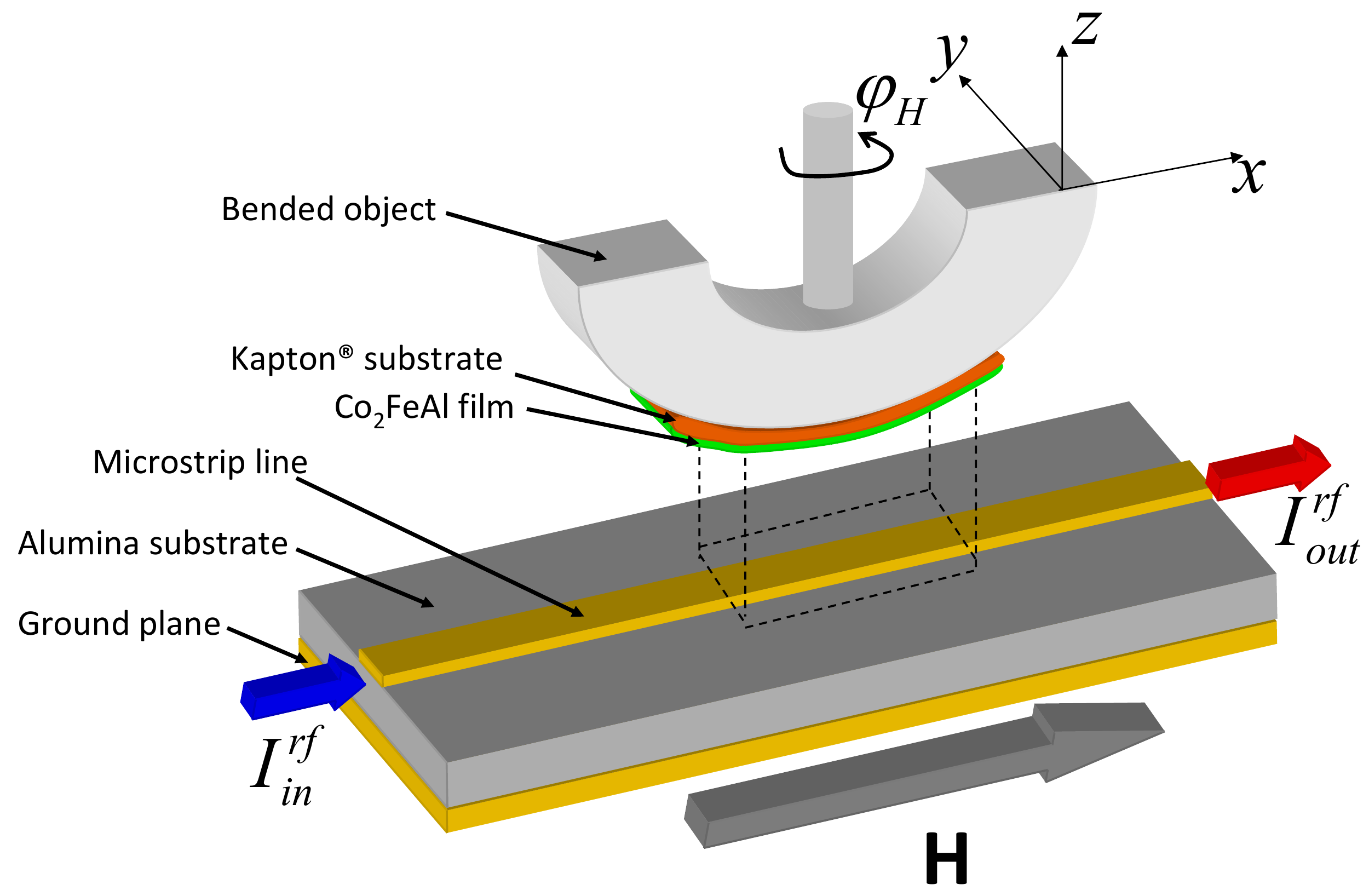}

\caption{Sketch of the microstripline resonator allowing for the resonance
field detection of the bended CFA film deposited onto flexible substrate.
$I_{in}^{rf}$and $I_{out}^{rf}$correspond to the injected and transmitted
radio frequency current (fixed at 10 GHz thereafter). The static magnetic
field $\vec{H}$ is applied along the microstripline.}

\label{FMR_sketch}
\end{figure}

The bending strain effect has been experimentally studied by microstripline
ferromagnetic resonance (MS-FMR), shown in Fig.\ref{FMR_sketch},
through uniform precession mode resonance field. Indeed, the resonance
field of the uniform precession mode is influenced by the magnetoelastic
behavior of the thin film. All the experimental MS-FMR spectra analyzed
in this work have been performed at room temperature at a fixed driving
frequency of 10 GHz. In order to quantitatively study the magnetoelastic
behavior of the thin film, we have analytically modeled the bending
strain effect on the ferromagnetic resonance field through a magnetoelastic
density of energy $F_{me}$ : 
\begin{equation}
F_{me}=-\frac{3}{2}\lambda\left(\gamma_{x}^{2}-\frac{1}{3}\right)\sigma_{xx}
\end{equation}

$\sigma_{xx}$ being the uniaxial stress due to bending while $\gamma_{x}$
correspond to the direction cosines of the in-plane magnetization.
$\lambda$ is the effective magnetostriction coefficient of the CFA
film. The relation between the principal stress component ($\sigma_{xx}$)
and the radius curvature $R$ is given by the following equation available
when the film thickness is very small as compared to the substrate
ones: 

\begin{equation}
\sigma_{xx}=E\frac{t}{2R}
\end{equation}

where $t$ is the whole sample thickness ($\sim$ the substrate thickness
in our case) and $E$ is the Young's modulus.

In these conditions, the resonance field of the uniform precession
mode evaluated at the equilibrium is obtained from the total magnetic
energy density $F$ as follows: 
\begin{equation}
\left(\frac{2\pi f}{\gamma}\right)^{2}=\left(\frac{1}{M_{s}\sin\theta_{M}}\right)^{2}\left(\frac{\partial^{2}F}{\partial\theta_{M}^{2}}\frac{\partial^{2}F}{\partial\varphi_{M}^{2}}-\left(\frac{\partial^{2}F}{\partial\theta_{M}\varphi_{M}}\right)^{2}\right)\label{eq:Smit_Beljers}
\end{equation}

In the above expression, $f$ is the microwave driving frequency,
$\gamma$ is the gyromagnetic factor ($\gamma=g\times8.794\times10^{6}$
s$^{-1}$ Oe$^{-1}$) while $\theta_{M}$ and $\varphi_{M}$ stand
for the polar and the azimuthal angles of the magnetization. It should
be noted here that the saturation magnetization ($M_{s}$) has been
measured by vibrating sample magnetometry ($M_{s}\simeq820$ emu.cm$^{-3}$).
The magnetic energy density $F$ is the sum of several contributions
including the Zeeman $F_{zee}$, the dipolar $F_{dip}$ and the magnetoelastic
$F_{me}$ and the out-of-plane magnetic anisotropy $F_{perp}=-K_{perp}\cos^{2}\theta_{M}$
(where $K_{perp}$ is the out-of-plane anisotropy constant) contributions.
Thereafter, $\varphi_{H}$ will correspond to the angle between the
in-plane applied magnetic field and the bending axis ($x$ direction)
as presented in Figure \ref{FMR_sketch}. In addition, an initial
in-plane uniaxial anisotropy (measured on the unbended sample) has
been put into evidence and is attributed to a non-equibiaxial residual
stress inside the magnetostrictive film induced by a slight initial
curvature of the sample after (or during) deposition. Thus, a magnetoelastic
energy term ($F_{me}^{residual}$ ) will be added to take into account
this initial anisotropy: 
\begin{equation}
F_{me}^{residual}=-\frac{3}{2}\lambda\Big(\big(\gamma_{x}^{2}-\frac{1}{3}\big)\sigma_{xx}^{residual}+\big(\gamma_{y}^{2}-\frac{1}{3}\big)\sigma_{yy}^{residual}\Big)
\end{equation}

$\sigma_{xx}^{residual}$ and $\sigma_{yy}^{residual}$ being the
in-plane principal residual stress tensor components and $\varphi_{resi}$
is the angle between $x$ axis and the slight initial curvature. In
these conditions, the resonance field can be extracted from the following
expression: $f^{2}=\left(\frac{\gamma}{2\pi}\right)^{2}H_{1}H_{2}$
where:

\begin{multline}
H_{1}=\\
4\pi M_{s}-\frac{2K_{perp}}{M_{s}}+H_{res}\cos(\varphi_{M}-\varphi_{H})+\frac{3\lambda}{M_{s}}\left(\sigma_{xx}\cos^{2}\varphi_{M}\right)+\\
\frac{3\lambda}{M_{s}}\Big(\sigma_{xx}^{residual}\cos^{2}(\varphi_{M}-\varphi_{resi})+\sigma_{yy}^{residual}\sin^{2}(\varphi_{M}-\varphi_{resi})\Big)\label{eq:H1}
\end{multline}

\begin{multline}
H_{2}=\frac{3\lambda}{M_{s}}\sigma_{xx}\cos2\varphi_{M}+H_{res}\cos(\varphi_{M}-\varphi_{H})+\\
\frac{3\lambda}{M_{s}}\left(\sigma_{xx}^{residual}-\sigma_{yy}^{residual}\right)\cos2(\varphi_{M}-\varphi_{resi})\label{eq:H2}
\end{multline}

In this formalism, and because $K_{perp}$ and $\gamma$ can be completely
determined at zero applied stress \cite{Zighem2010}, the main unknown
is the magnetostriction coefficient $\lambda$ since CFA single-crystal
elastic constants can be found elsewhere ($C_{11}$ = 253 GPa, $C_{12}$
= 165 GPa, $C_{44}$ = 153 GPa \cite{Gabor2011}). Indeed, in these
conditions, the Young's modulus of a polycrystalline film (our case
since it is deposited onto polymer substrate) can be estimated using
suitable averaging (homogenization method detailed by Faurie \textit{et
al.} \cite{Faurie2010}), requiring the knowledge of the grain orientations
distribution developed during film deposition. 

The 20nm-thick CFA film was grown on Kapton\textregistered{} substrate
(of thickness 127.5 \textmu{}m $\sim t$) using a magnetron sputtering
system with a base pressure lower than $3\times10{}^{\text{\textminus}9}$
Torr. CFA thin film was deposited at room temperature by dc sputtering
under an Argon pressure of $1\times10{}^{\text{\textminus}3}$ Torr,
at a rate of 0.1 nm.s$^{-1}$. The CFA film were then capped with
a Ta (5 nm) layer. Finally the CFA film is mounted on curved Aluminum
blocks of different radii after the characterization of the CFA thin
film (unbended). Indeed the stacking film is widely thinner that the
substrate (more than three order of magnitude) so that the uniaxial
stress $\sigma_{xx}$ can be considered as homogeneous in the film
thickness. X-ray diffraction measurements showed that no preferential
orientation developed during film growth. Being given this random
grain orientation distribution, we can estimate the Young's modulus
to be $E=243\times10^{10}$ dyn.cm$^{-2}$ ($\equiv243$ GPa). 

\begin{figure}
\includegraphics[clip,width=8.5cm]{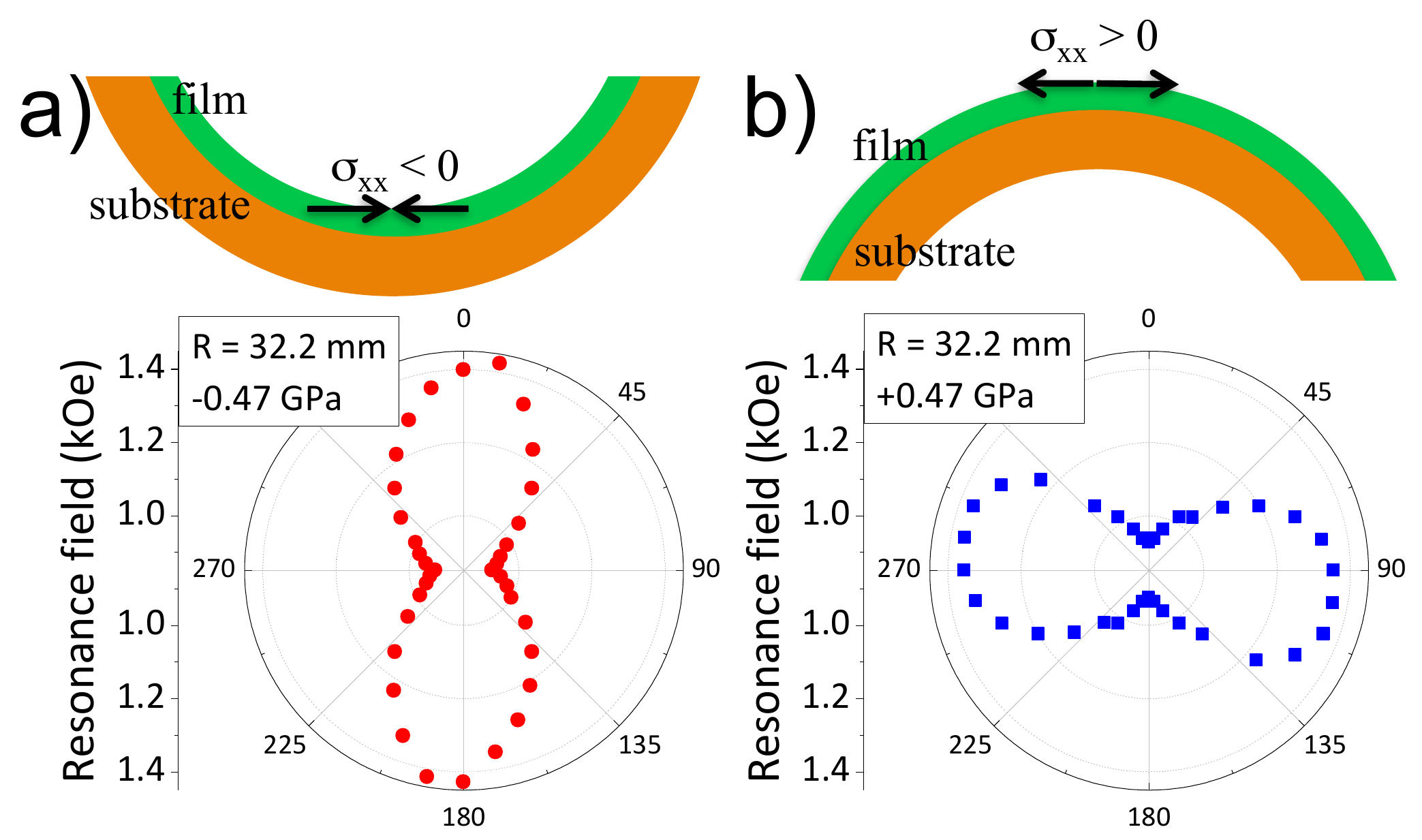}

\caption{Azimuthal angle dependance of the resonance field in polar representation
with the corresponding bending effect on the stress state (either
negative (a) or positive (b))for 20 nm thick CFA film placed on aluminum
block with 32.2 mm radius.}

\label{Bending_Effect}
\end{figure}

The thin film has been placed on small pieces of circular Aluminum
blocks of known radii $R$ (13.2 mm, 32.2 mm, 59.2 mm and infinite
(flat surface)) and analyzed by MS-FMR at 10 GHz driven frequency.
In our conditions, these radii values correspond respectively to the
following values of applied stress $\sigma_{xx}$ : 1.15 GPa, 0.47
GPa, 0.26 GPa, 0 GPa. Moreover we have stressed the thin film compressively
(Fig.\ref{Bending_Effect}-a) and tensily (Fig.\ref{Bending_Effect}-b)
so that we have studied three opposite stress states and the zero
stress state (unbended sample). We can see in Fig.\ref{Bending_Effect}
($R$ = 32.2 mm) that the sign change for $\sigma_{xx}$ in the thin
film induces a switching of the uniaxial anisotropy easy axis as revealed
by the angular dependance of the resonance field for the two opposite
stress values $\sigma_{xx}=-0.47$ GPa and $\sigma_{xx}=+0.47$ GPa.
In our configuration, the $x$ axis corresponds to $\varphi_{H}=90\text{\textdegree\ and 270\textdegree}$
and $y$ axis corresponds to $\varphi_{H}=0\text{\textdegree\ and 180\textdegree}$
. We will see that the apparent slight misalignment between the easy
axis and the $x$ axis in Fig.\ref{Bending_Effect}-a ($\sigma_{xx}<0$)
and the $y$ axis in Fig.\ref{Bending_Effect}-b ($\sigma_{xx}<0$)
respectively is mainly due to an initial uniaxial anisotropy in the
thin film (before applied bending) at about 30\textdegree{} from the
$x$ axis.

\begin{figure}
\includegraphics[clip,width=8.8cm]{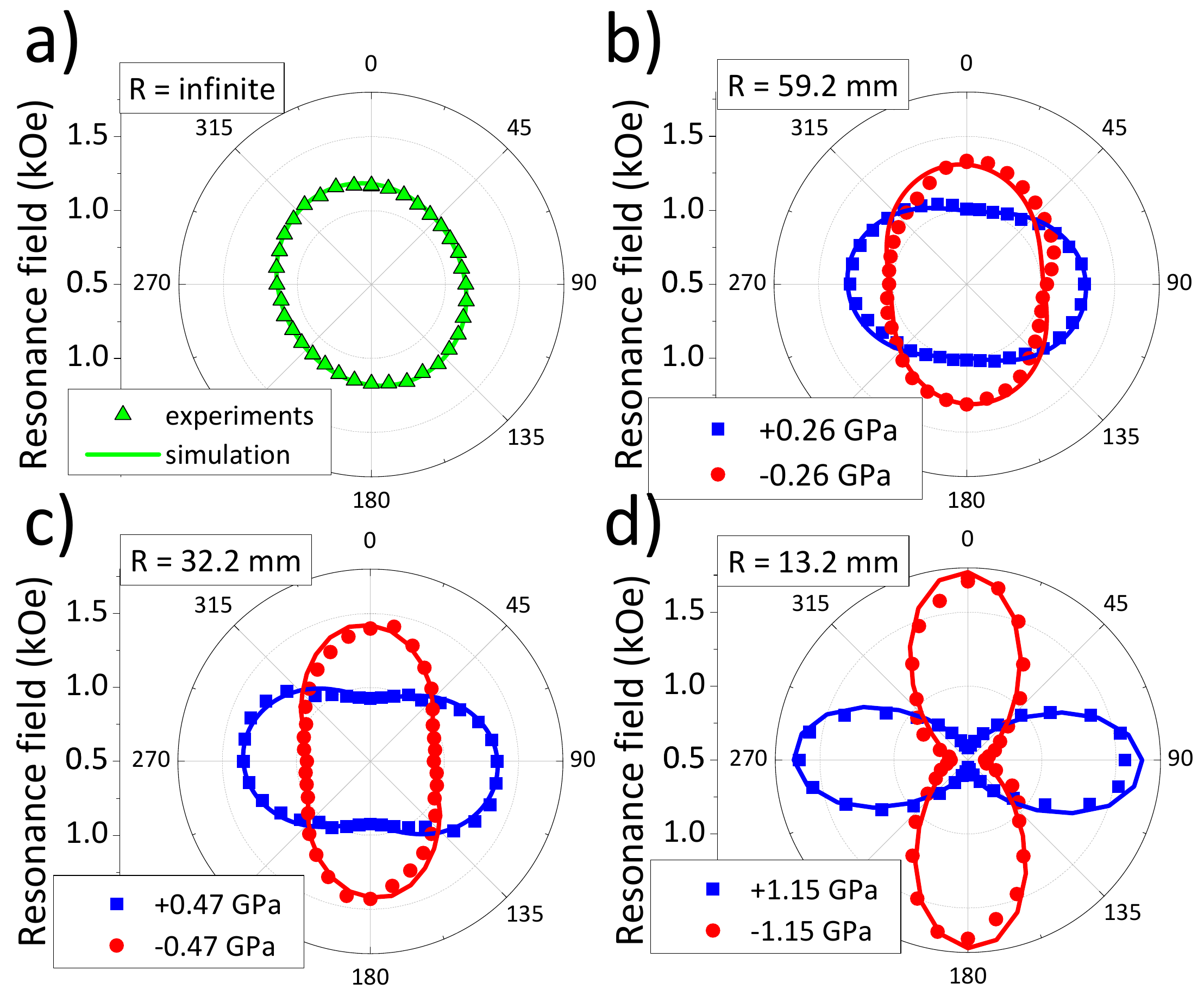}

\caption{Angular dependance of the resonance field (at 10 GHz) for $R=\infty$
(a), $R=59.2$ mm (b), $R=32.2$ mm (c), $R=13.2$ mm (d). In figures
(b), (c) and (d) are shown the opposite stress states.}

\label{Angular_Dependences}
\end{figure}

Fig.\ref{Angular_Dependences}-a shows the angular dependance of the
resonance field for unbended sample. This initial uniaxial anisotropy,
which has been observed in previous work in magnetic thin films deposited
on flexible substrates \cite{Zhang2013,Zighem2013}, is generally
attributed to a slight initial unavoidable curvature of the sample
after deposition when using such substrates. In Fig. \ref{Angular_Dependences}-b,
\ref{Angular_Dependences}-c and \ref{Angular_Dependences}-d are
shown the angular dependencies of the resonance fields for all the
applied stress states. Full symbols show the experimental data while
continuous lines show the fit of the data using the formalism detailed
below. Being given a saturation magnetization of 820 emu.cm$^{-3}$,
the best fits to the whole angular dependencies (performed at $f=10$
GHz) allowed for the determination of the following parameters: $\left|\sigma_{xx}^{residual}-\sigma_{yy}^{residual}\right|=90$
MPa, $\varphi_{resi}=30\text{\textdegree}$, $K_{perp}=53\times10^{4}$
erg.cm$^{-3}$, $\gamma=1.835\times10^{7}$ s$^{-1}$.Oe$^{-1}$ and
$\lambda=14\times10^{-6}$. The first two parameters characterize
the initial uniaxial anisotropy (magnetoelastic) which has an amplitude
of around 25 Oe slightly misaligned with respect to the $x$ axis
(see Fig.\ref{Angular_Dependences}-a). 

Interestingly, in-plane magnetostriction coefficient at saturation
$\lambda$ can be determined in a simplest way. Indeed, at $\varphi_{H}=0$,
the magnetization is aligned along the applied magnetic field ($\varphi_{H}\sim\varphi_{M}$)
and a linear resonance field-dependence of slope $\lambda$ is derived
as function of $\frac{-3\sigma_{xx}}{M_{s}}$: 
\begin{align}
H_{res}(\varphi_{H} & =0)=H_{3}-\frac{3\lambda}{M_{s}}\sigma_{xx}
\end{align}

\begin{figure}
\includegraphics[clip,width=8.5cm]{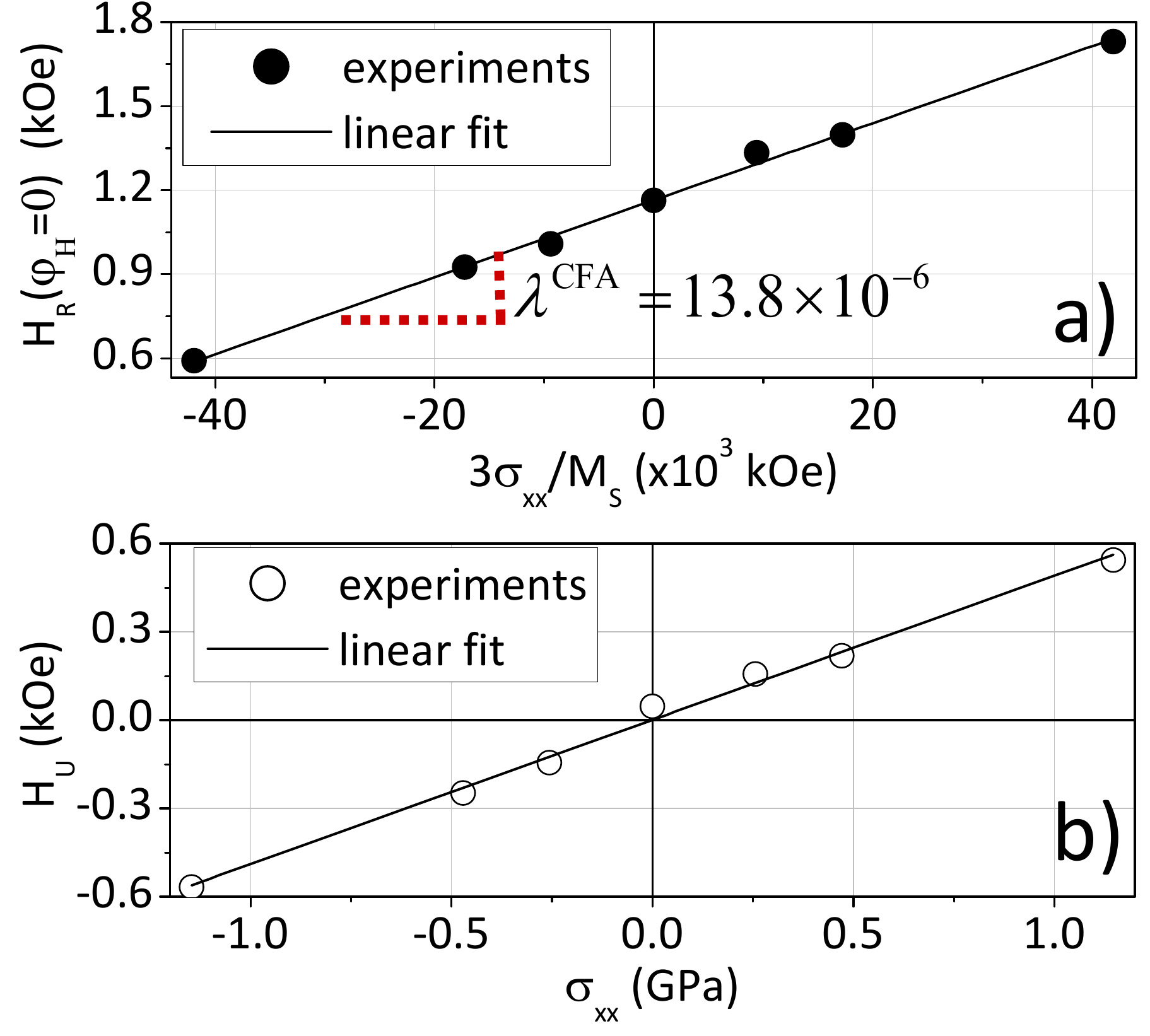}

\caption{(a) Resonance field as function of $\left(-\frac{\sigma_{xx}}{3M_{s}}\right)$.
Here the slope of the linear curve is the magnetostriction coefficient
$\lambda^{CFA}=13.8\times10^{-6}$. (b) Anisotropy field as function
of applied stress $\sigma_{xx}$ induced by bending.}

\label{Lambda_Hu_Fits}
\end{figure}

$H_{3}$ being a term that is independent of the applied stress $\sigma_{xx}$.
Thus, by plotting the experimental resonance field as function of
of $\frac{-3\sigma_{xx}}{M_{s}}$ at $\varphi_{H}=0$ (Fig.\ref{Lambda_Hu_Fits}-a),
a simple linear fit gives in more a direct way the value of $\lambda^{CFA}=13.8\times10^{-6}$.
It should be noted that this fitting procedure does not require any
knowledge on the residual stress state, the initial anisotropies or
the gyromagnetic factor. The value found here is in good correlation
with those found using the complete fit. This positive value means
that a uniaxial tensile stress along the $x$ axis will make easier
this axis for the magnetization direction. This effect is illustrated
in Fig. \ref{Lambda_Hu_Fits}-b that shows the linear dependance of
the effective bending-induced in-plane anisotropy field $H_{U}$ (where
$2H_{U}\simeq\left(H_{res}(\varphi_{H}=0)-H_{res}(\varphi_{H}=90\text{\textdegree})\right)$)
as function of the applied stress $\sigma_{xx}$. In our experimental
conditions, the extreme values of $H_{U}$ are roughly -0.6 kOe and
0.6 kOe that are very high being given the small effort to bend this
kind of flexible samples. Indeed, the anisotropy field induced by
bending would be enough to compensate ones already present in patterned
thin films showing sub-micronic lateral dimensions as encoutered in
spin valve sensors for instance\cite{Freitas2007}.

Finally, concerning the field of flexible spintronics, one difficulty
will be to get materials with large GMR (or TMR \cite{Barrault2012,Bedoya-Pinto2014})
remaining almost constant during external loading, \textit{i.e.} with
very small in-plane magnetostriction coefficient (less than $10^{-6}$).
This would be either intrinsic to the material (depending on the alloying
elements and stoichiometric) either due to microstructural features
such as crystallographic texture since the in-plane magnetostriction
coefficient of polycrystalline thin films depends on single-crystal
coefficients ($\lambda_{111}$ and $\lambda_{100}$ for cubic symmetry)
and on grains orientations distribution \cite{Bartok2011,Zighem2013}.

In conclusion, we have shown that the magnetization in CFA Heusler
alloy deposited on flexible substrate can be easily manipulated by
bending the sample. Obviously, a slight curvature of the sample induces
an uniaxial anisotropy that is generally present in such flexible
samples. Moreover, by modeling the bending strain effect, and by adjusting
the analytical model to the FMR data, it has been possible to extract
the in-plane magnetostriction coefficient: $\lambda^{CFA}=13.8\times10^{-6}$.
In order to be applied in GMR flexible systems, it is imperative to
deposit Heusler alloys with lower coefficient (at least ten times
lower), in order to keep a constant value of GMR if sample bending
occurs.
\begin{acknowledgments}
The authors gratefully acknowledge the CNRS for his financial support
through the ``PEPS INSIS'' program (FERROFLEX project) and by the
Université Paris 13 through a ``Bonus Qualité Recherche'' project
(MULTIDYN). Tarik Sadat (PhD student at Paris 13th university) is
thanked for helping us in programming our resonance field ``Mathematica-code''.
Authors would like to thank Frédéric Lombardini, engineer-assistant
at LSPM-CNRS, for circular blocks machining. M.S.G, T.P. and C.T.
acknowledge financial support through the Exploratory Research Project
\textquotedbl{}SPINTAIL\textquotedbl{} PN-II-ID-PCE-2012-4-0315.\end{acknowledgments}

\end{document}